\documentclass[twocolumn,notitlepage,aps,pra,superscriptaddress,amsmath,amssym,floatfix,longbibliography]{revtex4-1}
\usepackage{subcaption}
\usepackage{graphicx}
\usepackage{bm}
\usepackage{color}
\usepackage{amssymb}
\usepackage{enumerate}
\usepackage{comment}

\def\la{\langle}
\def\ra{\rangle}

\def\be{\begin{equation}}
\def\ee{\end{equation}}

\begin{document}

\newcommand{\bigjprob}{{\mathcal{P}}}
\newcommand{\bigprob}{_{\bm{q}_F}{\mathcal{P}}_{\bm{q}_I}}
\newcommand{\cum}[1]{\llangle #1 \rrangle}       					
\newcommand{\op}[1]{\hat{\bm #1}}                					
\newcommand{\vop}[1]{\vec{\bm #1}}
\newcommand{\opt}[1]{\hat{\tilde{\bm #1}}}
\newcommand{\vopt}[1]{\vec{\tilde{\bm #1}}}
\newcommand{\td}[1]{\tilde{ #1}}
\newcommand{\mean}[1]{\la#1\ra}                  					
\newcommand{\cmean}[2]{ { }_{#1}\mean{#2}}       				
\newcommand{\pssmean}[1]{ { }_{\bm{q}_F}\mean{#1}_{\bm{q}_I}}
\newcommand{\ket}[1]{\vert#1\ra}                 					
\newcommand{\bra}[1]{\la#1\vert}                 					
\newcommand{\ipr}[2]{\left\la#1\mid#2\right\ra}            				
\newcommand{\opr}[2]{\ket{#1}\bra{#2}}           					
\newcommand{\pr}[1]{\opr{#1}{#1}}                					
\newcommand{\Tr}[1]{\text{Tr}(#1)}               					
\newcommand{\Trd}[1]{\text{Tr}_d(#1)}            					
\newcommand{\Trs}[1]{\text{Tr}_s(#1)}            					
\newcommand{\intd}[1]{\int \! \mathrm{d}#1 \,}
\newcommand{\dd}{\mathrm{d}}
\newcommand{\fullint}{\iint \! \mathcal{D}\mathcal{D} \,}
\newcommand{\drv}[1]{\frac{\delta}{\delta #1}}
\newcommand{\partl}[3]{ \frac{\partial^{#3}#1}{ \partial #2^{#3}} }		
\newcommand{\smpartl}[4]{ \left( \frac{\partial^{#3} #1}{ \partial #2^{#3}} \right)_{#4}}
\newcommand{\smpartlmix}[4]{\left( \frac{\partial^2 #1}{\partial #2 \partial #3 } \right)_{#4}}
\newcommand{\limit}[2]{\underset{#1 \rightarrow #2}{\text{lim}} \;}
\newcommand{\funcd}[2]{\frac{\delta #1}{\delta #2}}
\newcommand{\funcdiva}[3]{\frac{\delta #1[#2]}{\delta #2 (#3)}}
\newcommand{\funcdivb}[4]{\frac{\delta #1 (#2(#3))}{\delta #2 (#4)}}
\newcommand{\funcdivc}[3]{\frac{\delta #1}{\delta #2(#3)}}
\definecolor{dgreen}{RGB}{30,130,30}

\title{Fundamental Limits on Subwavelength Range Resolution}

\author{Andrew N. Jordan}
\affiliation{Institute for Quantum Studies, Chapman University, Orange, CA 92866, USA}
\affiliation{Department of Physics and Astronomy, University of Rochester, Rochester, NY 14627, USA}
\email{jordan@chapman.edu}
\author{John C. Howell}
\affiliation{Institute for Quantum Studies, Chapman University, Orange, CA 92866, USA}
\affiliation{Racah Institute of Physics, The Hebrew University of Jerusalem, Jerusalem, Israel, 91904}
\date{\today}

\begin{abstract}
We establish fundamental bounds on subwavelength resolution for the radar ranging problem, ``super radar''.  Information theoretical metrics are applied to probe the resolution limits for the case of both direct electric field measurement and photon-counting measurements.  To establish fundamental limits, we begin with the simplest case of range resolution of two point targets from a metrology perspective. These information-based metrics establish fundamental bounds on both the minimal discrimination distance of two targets as well as the precision on the separation of two subwavelength resolved targets. For the minimal separation distance, both the direct field method and photon counting method show that the discriminability vanishes quadratically as the target separation goes to zero, and is proportional to the variance of the second derivative of the electromagnetic field profile.  Nevertheless, robust subwavelength estimation is possible. Several different band-limited function classes are introduced to optimize discrimination.  We discuss the application of maximum likelihood estimation to improve the range precision with optimal performance.  The general theory of multi-parameter estimation is analyzed, and a simple example of estimating both the separation and relative strength of the two point reflectors is presented.
\end{abstract}

\maketitle
\section{Introduction} \label{sec:intro}
Radar and lidar are powerful remote sensing techniques used in a wide array of applications, such as archaeology, utilities, navigation, topographical mapping, distance and velocity detection, civilian air traffic control, and tracking and detection of satellites and spacecraft \cite{center1997electronic,doviak2006doppler,wang2016overview,komissarov2019partially,francke2012review,neal2004ground,li2017review,dong2017lidar}.  Range resolution in radar \cite{levanon2004radar,neal2004ground,skolnik1980introduction} or lidar \cite{nuss2020frequency,behroozpour2017lidar,caldwell2022time} is the smallest measurable distance between two objects within the same angular direction.  Range resolution is vital in target characterization and imaging. For decades, disambiguation of the two objects was determined solely by temporally resolving the pulses which reflected from each of the objects, see e.g. \cite{sheriff1977limitations,neal2004ground}.  When the range resolution distance $d_r$ obeys the inequality
$d_r\geq c\tau/2$, where $c$ is the speed of light, $\tau$ is the pulse width and 2 comes from the round trip of the pulse \cite{skolnik1980introduction,levanon2004radar,zhou2019quantum}, the two objects are considered resolved. 

Recently, the authors showed that the coherent interference created from the pulses reflecting from the objects could be used to greatly improve upon existing definitions of range resolution \cite{PhysRevLett.131.053803}.  When two identical, but slightly shifted in time, radar or coherent lidar pulses are summed, they create a new composite pulse.  The novel interference technique employed by the authors relied on a self-referential signal that compared the resulting interference feature in a region of steep gradients with a zero gradient portion of the signal.  Single parameter estimation allowed for precise determination of small relative shifts between two component pulses achieving far superior resolution than the inverse bandwidth can achieve.  

The critical insight is to use an amplitude-based resolution paradigm together with parameter estimation, rather than a temporally-resolved paradigm.  The self-referencing nature of the estimation using only relative amplitudes within the returning pulse, rather than absolute amplitude permits us to distinguish target properties from loss. We assume here that the character of the pulse shape is preserved by the intermediate medium.  In general, a frequency-dependent index of refraction will cause pulse distortion that will need to be accounted for in realistic applications. However, in establishing fundamental bounds, we focus on the simplest case of an undistorted pulse shape.  We note a fundamentally different kind of parameter estimation applied to incoherent imaging permits the estimation of the distance between two point sources better than the Rayleigh criterion using mode sorting of spatial optical modes, see e.g.  \cite{tsang2009quantum,tsang2017subdiffraction,wadood2021experimental,bearne2021confocal}, which can be extended to multi-parameter estimation \cite{yang2019optimal,qi2022quantum}.

In the previous work \cite{PhysRevLett.131.053803}, we used a balanced self-referencing estimation method to compare ratios of differences of the measured function at different space (or time) points to define a signal that was used to infer the distance between the two objects.  While the basic features of the super-radar functions were outlined and demonstrated in our previous paper, only minimal discussion of noise was presented.  Here we more formally address fundamental range resolution bounds using information theoretic metrics. We give treatments of direct measurement of electric fields of pulses of electromagnetic radiation, which is most suitable for radar applications, as well as in the photon-counting case, that is especially useful for coherent lidar.  An important concept for the noise models we describe is the Fisher information of the separation parameter.  This information metric sets a bound on the smallest variance of the parameter for large data sets over all estimators in the large data set limit, which sets the fundamental bound of our method.  We also find that the overlap between waveforms of different separations as well as the quantum Fisher information are useful and give additional insight into the parameter estimation.  We discuss different estimators, including maximum likelihood as well as Bayesian estimation methods.  We end with a discussion of the generalization to multi-parameter estimation, that will be critical to make this technique useful in realistic applications.

The paper is organized as follows:
In Sec.~\ref{sec:range}, we introduce the basic concepts of the remote sensing range resolution task.  In Sec.~\ref{sec:field}, we consider direct electric field detection, as appropriate for radio or microwave frequency electromagnetic radiation and radar applications.  There we compute the Fisher information of the target distance, that sets the best-case precision of range resolution, relative to the returning signal amplitude to detector noise ratio.  In Sec.~\ref{sec:photon}, we consider a photon-counting detector as appropriate for optical frequencies and lidar applications, and consider three different information theoretic metrics: A wave overlap metric to distinguish different values of the unknown parameter, the classical Fisher information for spatially (or temporally) resolved photon detection, and the quantum Fisher information metric, where the measured basis is optimized to extract maximal information about the parameter.  In Sec.~\ref{sec:states} we discuss different kinds of waveforms that are able to resolve subwavelength structure using this technique.  We discuss band-limited pulses, a line-segment triangle wave, and frequency combs. Generalizations to multiparameter estimation as derived in Sec.~\ref{sec:mpe}, and a two-parameter example is discussed consisting of range resolution and relative reflection strength. We conclude in Sec.~\ref{sec:conc}.

\section{Range Resolution in Remote Sensing} \label{sec:range}

We consider the simplest case of two equal-amplitude point scatterers of electromagnetic radiation, and our task is to determine ($i$) the minimal discrimination distance of the two targets and ($ii$) the precision on the range between them, focusing on subwavelength range resolution.  We introduce a simple one dimensional model, where the electric field has a spatial envelope of the form $f(x)$ is sent out and a return wave of the form 
\be
f_l(x) = \frac{1}{2} (f(x-l/2) + f(x+l/2) ), \label{fl}
\ee
is detected.  For simplicity, we have chosen to set the origin halfway between the two scattering centers.  Here $l$ is the distance between the two scatterers that we wish to estimate.
In a remote sensing context, the amplitude of the returning pulse is typically attenuated from the outgoing intensity by many orders of magnitude, so the absolute amplitude of the pulse is assumed to be unrelated to the target properties, and cannot not be used in the estimation task.  Consequently, we must consider the {\it normalized state} (\ref{fl}) and consider either the amplitude of the returning field relative to the detector noise, or the number of {\it detected} photons as the metrological resource.  In the sections below we assume the function $f$ is analytic and normalized.  We are most interested in the case where the function is {\it band-limited}, having an upper frequency cut-off $\Omega$ and when the range resolution parameter $l < 2\pi/\Omega$, breaking the long-standing trade-off between target resolution and wave carrier frequency.  We note that we have set the velocity of the pulse inside the medium equal to unity.  

\section{Field Detection} \label{sec:field}
In this section, we focus on the detection of the electric field of the radiation via the use of amplifiers.  We consider primarily radio frequencies in this section for radar applications.
In this frequency range, direct detection of the field is possible via amplification of the detected electromagnetic signals, leading to noise added to the signal waveform that is related to the effective amplifier temperature.  Here we consider a model to determine the ultimate limits of the precision on the distance between two scattering centers.  Let the detected signal from the amplifier be $s(t)$, and we time-discretize in units of the time-resolution of the detector $\Delta t$.  We convert to an effective distance $x$ to connect to the above discussion via the propagation speed of the wave and add a noise term arising from the amplifier,
\be
s_i = {\cal N}_l f_l(x_i) + \sigma \xi(x_i), \label{signal}
\ee
where the signal has been rescaled so that the overlap signal is normalized with the normalization factor ${\cal N}_l$, which is $l$-dependent.  Since the noise power of the amplified signal is known from the detector properties, the total signal can be scaled appropriately to normalize it.  The normalization condition yields the following expression for ${\cal N}_l$:
\be
{\cal N}_l^2 = \frac{2}{1+\int_{-\infty}^\infty  dx f(x+l/2) f(x-l/2)}.
\ee
Clearly ${\cal N}_l \rightarrow 1$ as $l \rightarrow 0$ because we choose $f(x)$ to be normalized. 
The detected signal $s_i$ contains $\sigma$ which is the noise strength, $\Delta x$ which is the converted discretized space unit (using the wave speed) and $\xi(x_i)$ is a unit-strength, white noise random variable with correlation
\be
\la \xi(x_i) \xi(x_j) \ra = \delta_{ij}.
\ee
In a remote detection set-up, the returning signal strength is unknown, so we keep the signal normalized, and thus the noise level $\sigma$ of the detector will also be rescaled depending on the signal strength.

We can pen a probability model for the above signal + noise point of view as
\be
P(\{ s_i \}|l) = \prod_i \frac{1}{\sqrt{2 \pi \sigma^2}} \exp\left( - \frac{(s_i - {\cal N}_l f_l(x_i))^2}{2\sigma^2}\right), \label{prob-seq}
\ee
where $P_l(\{ s_i \}|l)$ is the probability density of the sequence of signal data points $s_i$, given the value of the parameter $l$.  The Fisher information \cite{fisher1925theory} about the parameter $l$ in the above distribution is defined as
\begin{eqnarray}
I(l) &=& \prod_i \int_{-\infty}^\infty \left(\frac{\partial \ln P_l(s_i)}{\partial l}\right)^2 P_l(s_i) ds_i \nonumber \\
&=& -\prod_i \int_{-\infty}^\infty \left(\frac{\partial^2 \ln P_l(s_i)}{\partial l^2}\right) P_l(s_i) ds_i.
\end{eqnarray}
Here $P_l(s_i)$ is the probability of a single data point given a fixed $l$.  The inverse Fisher information bounds the variance of any estimator $\hat l$ of the parameter $l$, for large data sets
\be
{\rm Var}[{\hat l}] \ge \frac{M}{I(l)},
\label{crbound}
\ee
where $M$ is the number of repetitions of the measurement, the Cram\'er-Rao bound \cite{cramer1999mathematical}. The Fisher information formalism has been widely applied to bound estimation precision in coherent optics, see e.g. \cite{jordan2014technical}.

For the super-radar ranging problem, the logarithm of the distribution turns the product into a sum, so the resulting field-detection Fisher information for $M=1$ is given by
\be
I_f = \frac{1}{\sigma^2} \sum_i \left(\frac{\partial ({\cal N}_l f_l(x_i))}{\partial l}\right)^2.
\ee
In white noise models, the total noise power grows as $\int dx_1 dx_2  \la \Delta s(x_1) \Delta s(x_1)\ra = \int dx \sigma^2 = \sigma^2 X = \sigma^2 \Delta x N$, where  $N$ is the number of data points.  Thus, we can account for making the time-bins (converted here to space bins) wider by defining the noise density $\Sigma^2 = \sigma^2 \Delta x$, so we can turn the sum into an integral to approximate the Fisher information as
\be
I_f \approx \frac{1}{\Sigma^2} \int dx \left(\frac{\partial ({\cal N}_l f_l(x))}{\partial l}\right)^2.  \label{fisher-l}
\ee
This result can be further simplified for the range resolution problem as
\be
\Sigma^2 I_f = -(\partial_l \ln {\cal N}_l)^2 + \frac{{\cal N}_l^2}{16} \int dx (f'(x+l/2) - f'(x-l/2))^2.  \label{fieldfi}
\ee
The small $l$ behavior can be found by expanding all terms to leading order in $l$.  The 0th and first order terms vanishes, leaving the quadratic approximation near $l=0$,
\be
\Sigma^2 I_f \approx \frac{l^2}{16} \left(  \int dx f''(x)^2 - \left(\int dx f'(x)^2 \right)^2 \right) + {\cal O}(l^3). \label{quad-approx}
\ee
Here we assume well-defined first and second derivatives of the function $f(x)$, as well as square integrability.

As we will explore later on in the paper, let us define the $f$-dependent quantity in the large parentheses as ${\rm var}[d^2]_f$, a measure of the variance of the second derivative of $f$, which is a measure of the variation of the function $f$ over its range in space (or time).
If we define $\delta[d^2]_f = \sqrt{{\rm var}[d^2]_f}$ as its standard deviation, then we can bound the relative error $\delta l/l$ of the range resolution $l$ in the deep subwavelength limit from (\ref{crbound}) as
\be
\frac{\delta l}{l} \ge \frac{4 \Sigma}{\sqrt{M} l^2 \delta[d^2]_f}.
\label{bound}
\ee
This is a main result of this paper.
We note that when $\delta[d^2]_f$ is comparable to $\Omega^2$, the band-limit of the pulse, implying $l^2 \delta[d^2]_f <1$, then by having many data sets with good signal-to-noise ratio (i.e. $\sqrt{M}/\Sigma \gg 1$), relative errors less than 1 are easily achieved.  Compared with the traditional criterion for range resolved targets, where $l$ must be larger than the pulse width, our bound (\ref{bound}) can beat this criterion by many orders of magnitude.

\subsection{Estimators}
\label{field-estimator}
We can implement several different estimators.  We first consider {\it maximum likelihood estimation}, or MaxLike, for short. We define the log-likelihood function ${\cal L}(l)$ as the natural logarithm of the probability of obtaining the sequence $\{ s_i \}$ of measurement outcomes given the value of the unknown parameter is $l$.  Maximum likelihood estimation is defined as assigning an estimator ${\hat l}_{ml}$ such that
\be
{\hat l}_{ml} = {\rm argmax} [{\cal L}(l)].
\ee
Here argmax is the value taken at the maximization of the log-likelihood function.  For our problem (\ref{prob-seq}), the log likelihood function is given by
\be
{\cal L} = - \sum_j \frac{(s_j - {\cal N}_l f_l(x_j))^2}{2 \sigma^2} + const. \label{llf}
\ee
If there are $M$ repetitions of the measurement for each $x$ value then the data is indexed as $s_j^{(k)}$, where $k$ ranges from 1 to $M$, and the sum in ${\cal L}$ is over both $j$ and $k$.
The MaxLike condition given the data set $\{ s_j^{(k)} \}$, for the estimated value of $l$ is given by the condition
\be
\partial_l \sum_{j,k}  s_j^{(k)} {\cal N}_l f_l(x_j) =0,
\ee
where we used the fact that the function is normalized at this signal to noise level.  
The above simple result is very appealing:  in practice, one can define two vectors ${\bf u,v}$ of size $N$ with elements consisting of the measured data for the first vector $u_j = s_j$, and the second is a series of trial vectors defined by elements $v_j = {\cal N}_l f_l(x_j)$ for a variety of different values of $l$.  The first vector can carry out the sum the data $u_j \rightarrow \sum_{k=1}^M s_j^{(k)}$ for different values of $M$ to gain improved accuracy.  Accuracy is improved because the signal grows as $M$ but the noise only grows as $\sqrt{M}$.  The dot product between the two vectors ${\bf u} \cdot {\bf v}$ is then plotted versus the different trial values of $l$.  When the dot product reaches its maximum, that is the MaxLike estimated value for $l$.

There is an intuitive reason why the above procedure is optimal.  Let us recall the starting point of this analysis (\ref{signal}).  Let $l_t$ be the true value of $l$ that is sampled with the data, and let $l$ be the trial value that will be scanned.  We can write the function to be maximized as
\be
\sum_j s_j {\cal N}_l f_l(x_j) \approx \int dx [{\cal N}_{l_t} f_{l_t}(x) + \xi(x)] {\cal N}_l f_l(x).
\ee
when integrating over the random variable, that term tends to be zero, while the first term is simply the overlap between two normalized vectors,
\be
\la  f_{l_t} | f_{l} \ra = 
{\cal N}_{l_t}  {\cal N}_l \int dx  f_{l_t}(x)  f_l(x), \label{overlap-general}
\ee
where we introduced a Dirac notation for the inner product of the two normalized waves.
The overlap can at most be 1 when the value $l = l_t$, which is where the maximum value is reached.  Any other value of $l$ must necessarily lead to an overlap less than one.  From this point of view, we see that even if the rescaling is not perfectly normalized, the log-likelihood function will still be maximized at the true value of $l$.

Indeed, we can go further and approximate the log-likelihood function (\ref{llf}) in the vicinity of the true value of $l$ by writing ${\cal N}_{l} f_{l}(x) \approx {\cal N}_{l_t} f_{l_t}(x) + \partial_l ({\cal N}_{l} f_{l}(x))(l-l_t)$ and taking an average over the fluctuations to find
\be
{\cal L} \approx const - \frac{M}{2 \Sigma^2} \int dx  (\partial_l {\cal N}_{l_t} f_{l_t}(x))^2  (l - l_t)^2.
\ee
Therefore the uncertainty of the MaxLike method is set by the width of the log-likelihood curve in the large data limit,
\be
\la {\cal L}(l) \ra = const - \frac{(l - l_{t})^2}{2 \sigma_{ml}^2},
\ee
where the statistical uncertainty on $l$ is given by $\sigma_{ml}$, which is Gaussian in the large data limit by the central limit theorem.  The variance is given by the inverse Fisher information,
\be
\sigma_{ml}^2 = \frac{1}{M I_f(l)}.
\ee
Without averaging, the log-likelihood is given for a single realization of the data processing as
\be
{\cal L}(l)  = const - \frac{(l - {\hat l}_{ml})^2}{2 \sigma_{ml,1}^2},
\ee
where $\sigma_{ml,1}^2$ is the $M=1$ variance. The statistic ${\hat l}_{ml}$ can be found by fitting data to a parabola, and further averaging over $M$ repetitions to improve precision.  If ${\hat l}_{ml,j}$ is the MaxLike estimator for the $j$th trial of the $M$ repetitions, then the variance $(1/M)\sum_j ({\hat l}_{ml,j} - \la {\hat l}_{ml}\ra)^2$ is given by
\be
{\rm Var}[{\hat l}_{ml}] = \frac{ {\rm Var}[{\hat l}_{ml, 1}]}{M} \ge {\frac{1}{M I_{f,1}}}, 
\ee
where $I_{f,1}$ is the one trial Fisher information. We assume each trial is independent of the others.  Formal proofs of the results above may be found in the statistics literature \cite{fisher1925theory,kay1993fundamentals,ly2017tutorial} and arise from the central limit theorem.

Another estimation technique is {\it Bayesian estimation}.  In this method a probability distribution is given to $l$ that quantifies our degree of belief about the value of $l$.  The prior distribution is defined as $P_0(l)$.  The likelihood is the probability distribution of the data, given an assumed (and fixed) value of $l$.  We now use Bayes theorem to find the probability of the value $l$, given a set of data taken from the above model,
\be
P(l| \{ s_j \}) = P( \{ s_j \} | l) P_0(l) / P(\{s_j \}),
\ee
where $P(\{s_j \}) = \int dl P( \{ s_j \} | l) P_0(l)$.  The new probability distribution $P(l| \{ s_j \})$ is the posterior distribution for $l$, given the data collected.  If the posterior distribution is maximized, then we see the Bayesian estimator can be shifted away from the maximum likelihood estimator based on prior knowledge.  However, if a uniform prior is chosen, then the maximum of the distribution coincides with the MaxLike estimator.  A more refined Bayesian estimator 
is given by the average of $l$ in the posterior distribution, and the variance of the estimator can be computed as the variance of $l$ in the posterior distribution.  This procedure is generally more resource intensive because an entire distribution needs to be calculated for each value of $l$.  While a uniform prior is common and leads to the MaxLike estimator, it may be advantageous to choose the Jeffreys prior because the Fisher information vanishes at $l=0$,
\be
P_0(l)= \frac{\sqrt{I_f(l)}}{n}, \quad n = \int dl \sqrt{I_f(l)}.
\ee
For this choice, $P_0(l)$ vanishes linearly near the origin as $P_0(l) \sim (l/4) \delta(d^2)$, where $\delta(d^2)$ is the standard deviation of the second derivative.

\section{Photon detection}
\label{sec:photon}
In this section, we consider the complimentary problem of using direct photon detection of radiation returning from a target.  This analysis is suitable for lidar applications.  The photons are assumed to be time-resolved and tagged with their arrival time $t_i$.  In the analysis below, we convert this to a position label $x_i$ with the wave speed.  We will consider three different estimation approaches to determine the range parameter $l$:  an inner product metric (overlap criterion), the Fisher information of the time-resolved photo-detected events, and finally, the quantum Fisher information, that optimizes over all possible choices of measurement basis.

\subsection{Overlap Metric - State Distinguishability}
The ability to distinguish the original waveform (with $l=0$) from a returned signal with non-zero $l$ can be quantified with the overlap of the two waveforms.  Perfect distinguishability is obtained when the overlap is zero, but statistical discrimination for the hypothesis testing is possible when the overlap is less than one, see e.g. \cite{jordan2015heisenberg}.
Let us define the overlap between the return signal and the outgoing signal (here we assume both are normalized, with ${\cal N}_l$ being the normalization of $|f_l\ra$) as
\begin{eqnarray}
\la f |f_l \ra &=& {\cal N}_l \int_{-\infty}^\infty \label{overlap}
dx f(x)^\ast f_l(x) \\
&=& {\cal N}_l  \int_{-\infty}^\infty
dx f(x-l/4) f(x+l/4).
\end{eqnarray}
Here we symmetrized and combined the two terms, taking $f$ real for simplicity.  we note that this overlap is a special case of (\ref{overlap-general}), focusing on the case where the task is to disseminate the separation distance of $l$ from the null hypothesis that there is no separation.

We can expand in powers of $l$ (assuming a differentiable function) and integrate by parts to find that while both the normalization and the overlap integral have an order $l^2$ term, in the product that cancel out, leaving an order $l^4$ term as the leading order behavior,
\begin{eqnarray}
{\cal N}_l &=& 1 + \frac{l^2}{8} \int_{-\infty}^\infty f'^2 dx \\
&+& \frac{l^4}{32} \left(\frac{3}{4} \left(\int_{-\infty}^\infty f'^2 dx\right)^2 - \frac{1}{3} \int_{-\infty}^\infty  dx f^{(4)} f\right) + \ldots, \nonumber
\end{eqnarray}
while
\begin{eqnarray}
\int_{-\infty}^\infty
&dx& f(x-l/4) f(x+l/4) = 1 - \frac{l^2}{8} \int_{-\infty}^\infty f'^2 dx \nonumber \\
&+& \frac{l^4}{384} \int_{-\infty}^\infty  dx f^{(4)} f + \ldots,
\end{eqnarray}
so the combined expansion is given by
\be
\la f |f_l \ra = 1 - \frac{l^4}{128} \left(  \int_{-\infty}^\infty  dx f^{(4)} f - \left(\int_{-\infty}^\infty f'^2 dx\right)^2\right)
+
\ldots  \label{overlap-eq}
\ee
Thus, in order to maximize distinguishability, we should minimize the above expression.  This is clearly in the special case where $l$ is small on the scale of the features of $f(x)$.

\subsection{Fisher Information for Position Measurements}
We now investigate the Fisher information metric.  We assume the system is measured in a time-resolved manner, that we convert to a position basis via the wave speed.
The Fisher information about the parameter $l$ for photon detection is defined for probability distributions $p_l(x)$ as
\begin{eqnarray}
I_p &=& \int_{-\infty}^\infty \left(\frac{\partial \ln p_l(x)}{\partial l}\right)^2 p_l(x) dx\\
&=& -\int_{-\infty}^\infty \left(\frac{\partial^2 \ln p_l(x)}{\partial l^2}\right) p_l(x) dx.
\end{eqnarray}
We can simplify this formula, if we consider the square of the photon state as our probability distribution
\be
p_l(x) = {\cal N}_l^2| f_l(x)|^2.
\ee
Here we have renormalized the distribution with a prefactor ${\cal N}_l$, such that the distribution $p_l(x)$ is normalized for all $l$.  

An important aspect of the radar ranging problem is that in remote sensing, the amount of radiation received from the target is typically many orders of magnitude reduced in intensity.  Therefore, the total Fisher information is the normalized probability of return times the total number of photons in the return signal.  
Inserting this formula for the radar ranging problem, we find the result $I_p = I_1+I_2$, where
\begin{eqnarray}
I_1 &=& \frac{{\cal N}_l^2}{4} \int_{-\infty}^{\infty} dx \left ( f'(x-l/2)^2 + f'(x+l/2)^2\right) \nonumber \\
&=&  \frac{{\cal N}_l^2}{2} \int_{-\infty}^{\infty} dx f'(x)^2.
\label{i1}
\end{eqnarray}
where the prime denotes an $x$ derivative and in the second equality, we shifted the integration variable.
\be
I_2 = -\frac{\partial^2}{\partial l^2} \ln {\cal N}_l^2. \label{i2}
\ee
We observe that the only $l$ dependence in the Fisher information occurs in the normalization of the function. Despite the different looking expressions for the photon-detection Fisher information $I_p$ of Eq.~(\ref{i1},\ref{i2}) and the field Fisher information Eq.~(\ref{fieldfi}), they are in fact equivalent up to a factor of 4, i.e. $4 \Sigma^2 I_f = I_p$.

The leading order behavior for small $l$ is given by expanding both $I_1$ and $I_2$ to second order in $l$ to find,
\begin{eqnarray}
I_2 &=& -\frac{1}{2} \int_{-\infty}^\infty dx f'^2  \\
&+& \frac{l^2}{2} \left( -\frac{3}{4} \left(\int_{-\infty}^\infty dx f'^2\right)^2 + \frac{1}{2} \int_{-\infty}^\infty dx f f^{(4)} \right) + {\cal O}(l^4).\nonumber
\end{eqnarray}
Here, the $(4)$ indicates the 4th order derivative with respect to $x$.  For $I_2$ we find
\begin{eqnarray}
I_1 &=& \frac{1}{2} \int_{-\infty}^\infty dx f'^2 \nonumber \\
&+& \frac{l^2}{8} \left(\int_{-\infty}^\infty dx f'^2\right)^2  + {\cal O}(l^4).
\end{eqnarray}
Putting these results together to find the total Fisher information, we see that the zeroth order in $l$ term vanishes, leaving only the second order term, $I_{tot} = I_1 + I_2$,
\be
I_{tot}  \approx \frac{l^2}{4} \left( \int_{-\infty}^\infty dx f f^{(4)} - \left(\int_{-\infty}^\infty dx f'^2\right)^2 \right),
\ee
with a 4th order correction in $l$.
Thus, at $l=0$, there is no Fisher information to be had, and we need to go to the second order term in $l$ for small $l$. We notice that the result to this order can be expressed in terms of the variance of an operator ${\hat O} = -d^2/dx^2$ in the state $f$.  Indeed, we will see this more explicitly in the Quantum Fisher Information Section below.


\subsection{Quantum Fisher Information Metric}
We now investigate the quantum Fisher information metric.  Quantum Fisher information permits the further optimization of the parameter over all possible choices of measurement basis \cite{helstrom1969quantum,braunstein1994statistical}.  The quantum Fisher information (QFI) about the parameter $l$ is defined for pure states as 
\be
F = 4 (\la \partial_l f_l | \partial_l f_l \ra - | \la \partial_l f_l | f_l \ra|^2).\label{qfi}
\ee
The quantum Fisher information metric lower bounds the variance of any estimator $\tilde l$ of the parameter $l$, 
\be
{\rm Var}[{\tilde l}] \ge \frac{M}{F},
\ee
where $M$ is the number of measurements taken, that we consider to be the number of detected photons.  Thus, we wish to maximize the information acquired about the parameter $l$ in order to minimize the statistical uncertainty.

From a quantum mechanical perspective, we can write the radar ranging state $f_l(x)$ as 
\be
|f_l\ra = {\cal N}_l (e^{-i l {\hat p}/2} |f\ra + e^{i l {\hat p}/2} |f\ra)/2,
\ee
where $\hat p = -i\partial_x$ is the momentum operator, so its exponential shifts the original state $f$ by an amount $\pm l/2$.  Here we normalized the state from the start.  Thus, the shifted state can therefore be written as
\be
|f_l \ra = {\cal N}_l \cos(l {\hat p}/2) |f\ra.
\ee
The normalization can be found explicitly,
\be
{\cal N}_l^2 = \frac{1}{\la f|\cos^2(l {\hat p}/2) |f\ra}.
\ee
We note that the $l$-derivative is given by
\be
|\partial_l f_l \ra = - \frac{{\hat p}}{2} \sin (l {\hat p}/2) |f\ra + {\cal N}_l' \cos(l {\hat p}/2) |f\ra.
\ee
Inserting these result into the expression for the quantum Fisher information, we find the second term in (\ref{qfi}) is zero, while the first may be written as
\be
F = 4 \left( \frac{\la f | {\hat A}^2 |f\ra \la f | {\hat B}^2|f \ra - \la f | {\hat A}{\hat B}|f\ra^2}{ \la f| {\hat B}^2 | f\ra^2}
\right),
\ee
where we define the operators
\be
{\hat A} = \frac{ {\hat p}}{2} \sin(l {\hat p}/2), \qquad
{\hat B} =\cos(l {\hat p}/2).
\ee
 This is a general result for the radar ranging problem with two equal-weight point scatterers.  We note that by the Cauchy-Schwarz inequality for operators, that the quantum Fisher information $F$ must be positive.  This statement can be proved by considering the expectation value of the square of the operator $\la f|{\hat B}^2|f\ra {\hat A} - \la f|{\hat A} {\hat B}|f\ra {\hat B}$, in the state $|f\ra$, which must be positive.

We now focus on the special case where $l$ is small compared to the typical inverse momentum component of $f$, to obtain the leading order expression in $l$:
\be
F = \frac{l^2}{4}  {\rm Var}\left[ {\hat p}^2 \right]_f + {\cal O}(l^4),
\ee
where we defined the variance of the operator in the state $f$ as  ${\rm Var}[...]_f$.
We see immediately that in the position basis, we recover our previous result for the classical Fisher information, as is expected.

The variance can be calculated most easily in momentum space $(k)$ to find
\be
{\rm Var }\left[ {\hat p}^2 \right]_f = \int_{-\infty}^{\infty} dk |{\tilde f}(k)|^2 k^4 -  \left(\int_{-\infty}^{\infty} dk |{\tilde f}(k)|^2 k^2 \right)^2, \label{var-p}
\ee
where ${\tilde f}$ is the Fourier transform of $f$.

\begin{figure*}[thb]
\includegraphics[scale=1.1]{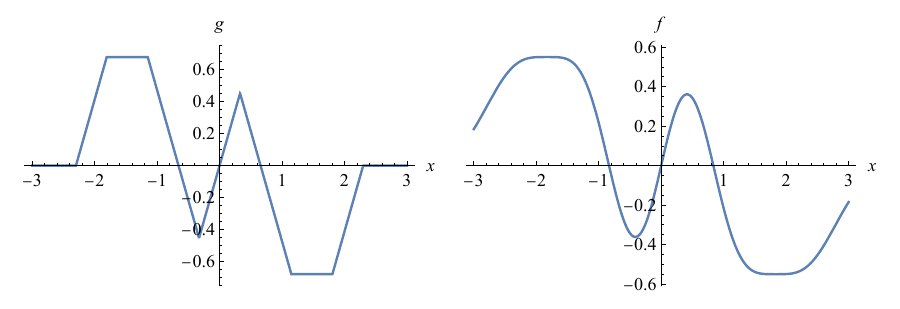}
\caption{Two waveforms for super range resolution.  The function $f$ is band-limited (see text), while the function $g(x)$ (left) is a line-segment ``triangle'' function that mimics it with segments of either steep slope or zero slope.  The triangle function is chosen so it has equal slopes and equal intervals.  Both functions are normalized and the slope of the triangle function $g$ chosen to match the slope of the band-limited function $f$ at $x=0$.}
\label{fig:funcs}
\end{figure*}

\section{Subwavelength Resolving States} \label{sec:states}
We now explore different states that could maximize the Fisher information.

\subsection{Sinc Derivatives}
We first consider the class of band-limited functions
\be
g_n(x) = N_n \frac{d^n}{dx^n} {\rm sinc}(k_0 x),
\ee
where $N_n$ is the normalization of the function and $k_0$ is the band limit we consider.  We consider $n$ even.  In this case, the functions are even, so we can write the overlap as a convolution $g_n \star g_n(-l/2)$. Using the convolution theorem, we can write the Fourier transform on the variable $l/2$ as
\be
\int_{\infty}^\infty d(l/2)
\la g |g_n \ra e^{ i k l/2} = |{\tilde g_n}(k)|^2,
\ee
where $\tilde g_n(k)$ is the Fourier transform.  In this case, the Fourier transform of the nth derivative of the sinc function is quite simple,
\be
{\tilde g}_n(k) = (ik)^n {\rm Box}[k-k_0]
\ee
where the Box function is 1 for $-k_0 < k < k_0$ and 0 elsewhere.  

We can calculate the variance (\ref{var-p}) for the class of function $g_n$ discussed above.  We find
\be
\la {\hat p}^2 \ra = \frac{2n+1}{2n+3} k_0^2, \quad \la {\hat p}^4 \ra =   \frac{2n+1}{2n+5} k_0^4,
\ee
Thus, the variance of ${\hat p}^2$ is given by
\be
{\rm Var}[{\hat p^2}] = k_0^4 \frac{4 (2n+1)}{(2n+5)(2n+3)^2}.
\ee
However, the Fisher information about $l$ vanishes in the large $n$ limit, so these functions are not helpful despite being band-limited.

\subsection{Frequency Combs}
We consider now delta-function peaks in the frequency space.  The simplest such combination is a monochromatic tone.  We consider the case often encountered in metrology of
\be
|{\tilde f}(k)|^2 = [\delta(k-k_0) + \delta(k+k_0)]/2.
\ee
This function gives a monochromatic tone with frequency $k_0$, say $\sin(k_0 x)$.  Unfortunately, while this function has good variance of ${\hat p}^2$, it has exactly zero variance of ${\hat p}^4$, so there is no information about the $l$ parameter.  This fact is clear from working in the original position (or time) space, where the function (\ref{fl}) is given by
\be
[\sin(k_0(x-l/2))  + \sin(k_0(x+l/2))]/2 = \sin(k_0 x) \cos(k_0 l/2),
\ee
so after normalization of the function, there is no $l$ information for any value of $l$.

However, it is not difficult to create a variance of ${\hat p}^2$ - we add in a DC offset at $k=0$, so that all are equally weighted,
\be
|{\tilde f}(k)|^2 = [\delta(k-k_0) + \delta(k) + \delta(k+k_0)]/3.
\ee
Then the variance is given by
\be
{\rm Var}[{\hat p}^2] = \frac{2 k_0^4}{9}.
\ee
More generally, we can consider a set of discrete frequencies $k_j$, with weights $c_j^2$, which then gives the results
\be
\la {\hat p}^2 \ra = \sum_j k_j^2 c_j^2, \quad 
\la {\hat p}^4 \ra = \sum_j k_j^4 c_j^2
\ee
For the case of a comb of $N$ regularly spaced frequencies $k_j = j k_0$ of weights $c_j^2$, we have
\be
|{\tilde f}(k)|^2 = \sum_{j=-N}^N c_j^2 \delta(k-j k_0).
\ee
If we take the $2N+1$ states to have the same weight $1/(2N+1)$, then we find the result
\be
{\rm Var}[{\hat p}^2] = \frac{k_0^4 N(N+1)(4N^2+4N-3)}{45} \approx 4 k_{max}^4/45,
\ee
where we denoted the highest frequency appearing as $k_{max} = N k_0$.

\subsection{Bandlimited Pulse}


We follow our earlier article Ref.~\cite{PhysRevLett.131.053803} and consider our interference class function to be a product of two functions, $f(x) = c_m(x) f_n(x)$, where $c_m$ is the ``canvas'' function, and $f_n$ is an approximation of the function we are interested in creating a band-limited version of \cite{vsoda2020efficient}. Beyond setting the band limits, the canvas function also enforces square integrability of $f(x)$ if chosen to decay sufficiently faster than the divergence of the polynomial $f_n(x)$, so we restrict to the case $m>n$.  We use the $\Omega$-bandlimited canvas functions
\begin{equation}
c_m(x)=\textrm{sinc}\left(\frac{\Omega x}{m}\right)^m
\end{equation}
where $m$ governs the power of decay.  The Fourier transform of $c_m(x)$ is strictly zero above frequency $\Omega$.  The polynomial $f_n(x)$ is defined as 
\begin{equation}
    f_n(x)=\sum_{k=0}^{n-1}a_k x^k,
\end{equation}
where $a_k$ are chosen to be the Taylor coefficients of the desired function $f$ up until order $n-1$.  We use the polynomial function
\begin{equation}
    f_n(x)=8x-14.3984x^3+4.77612x^5-0.82315x^7,
\end{equation}
along with the $c_{10}(x)$ canvas function with scaled dimensionless bandwidth $\Omega=2\pi$. 
A new method of creating band-limited functions that closely approximate a desired function in a finite region was recently introduced \cite{karmakar2023beyond}.
The function $f(x)$ has steep regions as well as flat regions to easily distinguish reflector range from loss, as shown in  the right panel of Fig.~\ref{fig:funcs} (we set the wave speed to one). 
A line segment function ``triangle function'' that captures the main features is shown in the left panel.  Both functions are normalized and the slope of the triangle function is matched to the slope of the band-limited function at $x=0$.

\begin{figure}[bht]
    \centering   \includegraphics[scale=0.8]{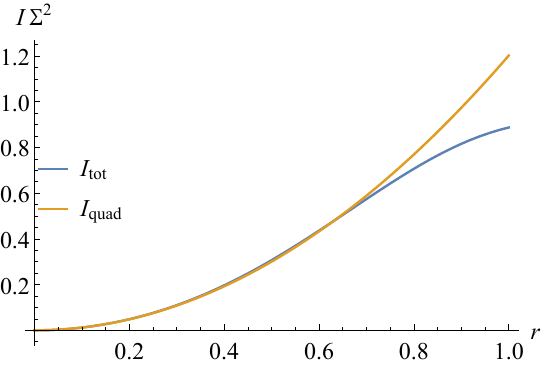}
    \caption{Fisher information times noise strength is plotted versus target separation, in units of $2\pi/\Omega$, in the case of field detection.  The entire parameter range shown is traditionally considered sub-resolved. The quadratic approximation $I_{quad}$ is compared with the exact result $I_f = I_{tot}$.}
    \label{fig:fieldfi}
\end{figure}

In Ref.~\cite{PhysRevLett.131.053803}, we used the estimator 
\begin{equation}
    S=\frac{A_{cmax}-A_{cmin}}{A_{lmax}-A_{lmin}}, \label{SignalEquation}
\end{equation}
where $A_{cmax}$, $A_{cmin}$ are the maximum and minimum amplitude of the function in the steep center region, respectively and $A_{lmax}$, $A_{lmin}$ are the maximum and minimum of the flat lobes of Fig.~\ref{fig:funcs}.  This estimator has many nice features, including its simplicity as well as resemblance to balanced detection \cite{dixon2009ultrasensitive}.  It is also linear in the parameter for much of the subwavelength resolved region.  Nevertheless, the MaxLike estimator in Sec.~\ref{field-estimator} is provably optimal.  This is because the MaxLike estimator incorporates the sensitivity of changes of the function across the entire function.  Nevertheless, the MaxLike estimator is nonlinear, and the simpler estimator above may be more practical in practice.  Nevertheless, we can quantify the advantages in precision the MaxLike estimator can provide.  Furthermore, given a known model or range of models, MaxLike can be generalized to cases where the simplifying assumptions of our two point scattering model break down.

The function $f(x)$ is used to compute the range resolution Fisher information in the field detection case (which is proportional to the Fisher information in the photon counting case, $I_p= 4 \Sigma^2 I_f$).  The scaled Fisher information $\Sigma^2 I_{f}$ is plotted in Fig.~\ref{fig:fieldfi} as a function of $r = l \Omega /2\pi$.  It vanishes as $l\rightarrow 0$, and behaves quadratically.  The quadratic approximation (\ref{quad-approx}) is also plotted as $I_{quad}$, calculating only the variance of the second derivative of the function $f(x)$, showing excellent agreement for small values of $l$ that are sub-resolved.

\begin{figure}[bht]
    \includegraphics[scale=0.8]{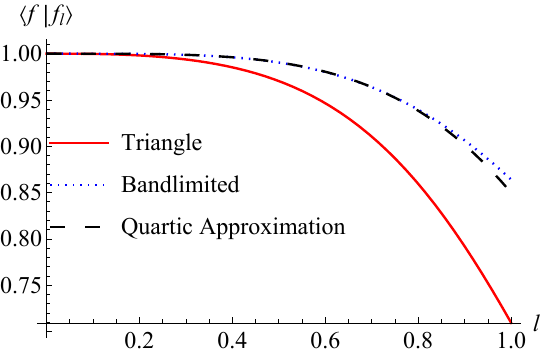}
    \caption{The inner product between the original function and its reflected version (\ref{overlap}) is plotted as a function of range separation $l$, in units $2\pi/\Omega$.  The overlap decays slowly as the 4th power of the separation, Eq.~(\ref{overlap-eq}), with the exact calculation shown in blue dots and the quartic approximation shown in black dashes.  The triangle function overlap, with matched derivative at the origin, is plotted in solid red.   The triangle overlap decays more quickly, indicating better resolvability, owing to its sharp cusps.  }
    \label{fig:overlap}
\end{figure}

The triangle function is non-analytic, so a Fisher information analysis is not suitable.  Nevertheless, we can compute the overlap metric in this case, and compare it to the band-limited function's behavior.  The plot of the overlap metric (\ref{overlap}) is shown as a function of the $l$ parameter in Fig.~\ref{fig:overlap} as a solid red curve.  Compared to the band-limited function discussed in this section (plotted in dotted blue), its non-analytic behavior leads to greater resolvability.  This is because as soon as there is any separation, the interfered function cuts the top off the triangle, leading to flat regions.  The drawback of this function is that it is no longer band-limited, and indeed contains high frequency content \cite{PhysRevLett.131.053803}. In practice the cusp of the triangle will be slightly smoothed.

\section{Multi-parameter Estimation}
\label{sec:mpe}
In the sections above we focused on single parameter estimation to set fundamental bounds on this technique.  However, the real power behind super radar will be to engage in many-parameter estimation for a variety of real world sensing tasks.    Therefore we give a brief treatment here of the fundamental limits of multi-parameter estimation, even if simpler estimation techniques are ultimately more practical.  

We define a collection of parameters $\theta_1, \theta_2, \ldots, \theta_p$ that we wish to estimate.  We focus on the field detection measurements here, so the detected signal is $s(x_i) = {\cal N}_{\vec \theta} f_{\vec \theta}(x_i)$, plus detector noise, where the $p$ parameters are encoded with a $p$-dimensional vector $\vec \theta$.  The statistical model is then given by (\ref{prob-seq}) where we replace the target range separation $l$ with $\vec \theta$, the set of parameters imprinted on the scattered wave from the collection of targets, which can now be quite general.  We define the single repetition Fisher information matrix $I_{jk}$ to be
\be
I_{jk} = -\int \prod_i ds_i  P(s_i | {\vec \theta}) \frac{\partial^2 \ln P( s_i |\vec \theta)}{\partial \theta_j \partial \theta_k}.
\ee
This matrix sets a bound of the covariance matrix of the parameters, ${\rm cov} [\theta_j \theta_k] = \langle \theta_j \theta_k \ra - \la \theta_j \ra \la  \theta_k \ra$, given by
\be
{\rm cov} [\theta_j \theta_k] \ge (I^{-1})_{jk}, \label{matrix-crb}
\ee
that generalizes the Cram\'er-Rao bound.  Here $I^{-1}$ is the matrix inverse of the Fisher information matrix.
An expression for the Fisher information matrix can be derived as a generalization to Eq.~(\ref{fisher-l}).  We find
\begin{eqnarray}
I_{jk} &=& \frac{1}{\Sigma^2} \int dx \frac{\partial ({\cal N}_{\vec \theta} f_{\vec \theta}(x))}{\partial \theta_j} 
\frac{ \partial ({\cal N}_{\vec \theta} f_{\vec \theta}(x))}{\partial \theta_k},\label{multifi}
\\
&=&  \frac{{\cal N}_{\vec \theta}^2}{\Sigma^2} \int dx \frac{\partial f_{\vec \theta}(x)}{\partial \theta_j} 
\frac{ \partial f_{\vec \theta}(x)}{\partial \theta_k} - \frac{1}{\Sigma^2 {\cal N}_{\vec \theta}^2}  \frac{\partial {\cal N}_{\vec \theta} }{\partial \theta_j} 
\frac{ \partial {\cal N}_{\vec \theta} }{\partial \theta_k}, \nonumber 
\end{eqnarray}
where we assume $f$ is analytic, bounded, and square integrable.  For $M$ repetitions of the measurement, a factor of $M$ appears as a prefactor to the Fisher information matrix. 
We note that optimal estimators saturating the matrix Cram\'er-Rao bound (\ref{matrix-crb}) are more difficult to find in the multi-parameter case and may not exist in general.

\subsection{Unequal Weight Reflectors}
As a simple example of multi-parameter estimation, we consider a slight generalization of the above two point scatterer model, where the weights of the two reflections are unequal.  In this case, the reflected function (\ref{fl}) becomes
\be
f_{l,p}(x) = p f(x-l/2) + (1-p) f(x+l/2),
\ee
where $p \in [0,1]$ is the relative weight of the two reflectors.  In this case, the Fisher information matrix is 2 by 2, where ${\vec \theta} = (l, p)$. The function normalization ${\cal N}_{l,p}$ depends on both parameters in general. Defining the matrix  $M_{jk} = \int dx \partial_{\theta_j} f_{\vec \theta}(x) \partial_{\theta_k} f_{\vec \theta}(x)$ that appears in Eq.~(\ref{multifi}) for convenience, we find
the matrix elements of $M$ to be
\be
M_{ll} = \int dx \left(\frac{\partial f_{l,p}}{\partial l}\right)^2,
\ee
where 
\be
\frac{\partial f_{l,p}}{\partial_{l}} = \frac{1}{2} \left[ -p f'(x-l/2) + (1-p) f'(x+l/2)\right].
\ee
The $pp$ element is
\be
M_{pp} =  \int dx \left(\frac{\partial f_{l,p}}{\partial p}\right)^2 = \int dx (f(x-l/2) - f(x+l/2))^2 .
\ee
For small $l$, we have the approximation $M_{pp} \approx l^2 \int dx f'(x)^2$.
The off-diagonal matrix element $lp$ is given by
\begin{eqnarray}
M_{lp} &=&  \int dx \frac{\partial f_{l,p}}{\partial l} \frac{\partial f_{l,p}}{\partial p} \nonumber \\
&=& (p-1/2) \int dx f(x+l/2) f'(x-l/2).
\end{eqnarray}
For small $l$ this element is approximated by $M_{lp} \approx (p-1/2) l \int dx f'(x)^2$.  The other off-diagonal matrix element $M_{pl} = M_{lp}$ by the equality of the mixed partial derivatives.  We notice the discrete symmetry $p \rightarrow 1-p$, $l \rightarrow -l$ both in the model and in the Fisher information matrix.

\section{Conclusions} \label{sec:conc}
We have explored the fundamental limits of remote sensing range resolution that is subwavelength, ``super-radar''.  By applying the formal methods of parameter estimation as applied to known waveforms reflected off two point targets with equal amplitude reflections (for simplicity), we quantified the smallest resolvable distance between the two targets and bounded the precision of range resolution as a function of separation.  We found general expressions for the Fisher information, which bounds the variance of the parameter of interest.  As the separation of the targets goes to zero, the information vanishes quadratically with a prefactor proportional to the variance of the second derivative, evaluated in the waveform.  We discussed different estimation methods, including Bayesian estimation and the optimal (nonlinear) maximum likelihood estimator, as well as a simpler balanced detection-type estimation method, that is linear for sufficiently large separation.  We also considered photon detection and derived both Fisher information for space (or time) resolved detection as well as quantum Fisher information, optimized to any measurement basis.  We find similar results there to the field detection: the information also vanishes quadratically as separation decreases, and the results are equivalent up to a constant of proportionality.  Nevertheless, these methods beat the traditional range resolution limits by orders of magnitude.  We explored different bandlimited waveforms that can be implemented in super-radar, optimizing the variance of the second derivative that will be continued in future work.  We also discussed multi-parameter estimation, and set fundamental bounds on that task as well.  In future work, we will implement these optimal estimators experimentally and demonstrate improved range resolution precision.  

\section{Acknowledgements}
Support from Chapman University and the Bill Hannon Foundation is gratefully acknowledged.  We thank Achim Kempf and Barbara \v{S}oda for helpful discussions.

\bibliography{refs}

\end{document}